\documentclass[a4paper,11pt]{article}

\usepackage{jheppub} % for details on the use of the package, please
\newcommand{\unl}[1]{\underline{#1}}

\newcommand{\msbar}{\overline{\text{MS}}}

\newcommand{\EQN}[1]{ \label{#1}}

\newcommand{\ovl}[1]{\overline{#1}}
\newcommand{\ice}[1]{\relax}

\newcommand{\beq}{\begin{equation}}
  
\newcommand{\eeq}{\end{equation}}
\newcommand{\re}[1]{(\ref{#1})}

\newcommand{\la}{\lambda}

\usepackage[safe]{textcomp}
\usepackage{lmodern} 
\usepackage{epsfig}
\usepackage{float}
\usepackage{amstext}  
\usepackage{amsfonts}
\usepackage{amsthm}
\usepackage{bm}       
\usepackage[bottom]{footmisc} 
\usepackage{array}                
\usepackage{xcolor} 
\usepackage{framed}
\usepackage{slashed}
\usepackage[absolute]{textpos}

\usepackage{multirow}

\newcommand{\p}{\partial}
\newcommand{\ddp}[1]{\frac{\partial}{\partial#1}}

\newcommand{\pmu}{\p_{\mu^2}}

\newcommand{\dmusq}[1]{\mu^2 \frac{\mathrm{d}#1}{\mathrm{d}\mu^2}}

\renewcommand{\dmusq}{\mu^2 \frac{\mathrm{d}}{\mathrm{d}\mu^2}}

\newcommand{\pmusq}{\mu^2 \frac{\p}{\p \mu^2}}

\newcommand{\dmu}{\mathrm{d}_{\mu^2}}

\newcommand{\fr}[2]{\frac{#1}{#2}}

\newcommand{\bea}{\begin{eqnarray}}
\newcommand{\eea}{\end{eqnarray}}
\newcommand{\be}{\begin{equation}}
\newcommand{\ee}{\end{equation}}
\newcommand{\ba}{\begin{align}}
\newcommand{\ea}{\end{align}}
\newcommand{\beas}{\begin{eqnarray}}
\newcommand{\eeas}{\end{eqnarray}}
\newcommand{\bes}{\begin{equation}}
\newcommand{\ees}{\end{equation}}
\newcommand{\bas}{\begin{align}}
\newcommand{\eas}{\end{align}}

\newcommand{\ssL}{{\mathcal L}}

\newcommand{\als}{\alpha_{\scriptscriptstyle{s}}}
\newcommand{\as}{a_{\scriptscriptstyle{s}}}
\renewcommand{\as}{a}
\newcommand{\lb}{\left(}
\newcommand{\rb}{\right)}

\definecolor{bluemar}{rgb}{0,0,.5}
\definecolor{redmar}{rgb}{.8,0,0}
\definecolor{greenmar}{rgb}{0,.5,0}

\title{
  A simple generalization of  the low-energy theorem for the
  effective Higgs-gluon-gluon coupling for the case of simultaneous
  decoupling of several heavy quarks}

\author{Konstantin G.\ Chetyrkin \ice{(\currenttime  \ \ \  \today)}  }

\affiliation{Institute for Nuclear Research of the Russian Academy of Sciences,
prospekt 60-letiya Oktyabrya 7a, Moscow 117312, Russia}

\emailAdd{chet4@ms2.inr.ac.ru}

\abstract{We extend in an extremely simple and straightforward way
  the well-known  low-energy theorem for
  an  effective Higgs-like scalar-gluon-gluon coupling \cite{Chetyrkin:1997un}
  (as well as a similar one  for  for the effective coupling of the Higgs-like  field to the
  light scalar quark currents) 
  in  QCD including arbitrary number of heavy quarks in addition to the light ones.
 The application of the generalized  low-energy theorem allows the extraction of the
  four-loop   effective Higgs-gluon-gluon coupling valid for extensions of the Standard Model with
  additional heavy quarks  from the 3-loop  results of  
  \cite{Grozin:2011nk} for the decoupling constant of $\alpha_s$.

}

\keywords{QCD, effective theories, Higgs production, multi-Higgs models}

\begin{document}
\maketitle
\flushbottom

\section{Introduction}

Due to a significant difference between the (double) top quark mass and the
Higgs boson mass the coupling of the Higgs boson to gluons is well described
by an effective theory obtained by integrating out the top quark field from
the full QCD Lagrangian.  A (QCD) Low Energy Theorem (LET) for the
Higgs-gluon-gluon coupling {induced by one or more quarks heavier than the
  Higgs mass} have been known since ages, originally for one and two loops
\cite{Shifman:1978zn,Shifman:1979eb,Voloshin:1985tc,Kniehl:1995tn,Kilian:1995tra,Inami:1982xt,Djouadi:1991tka}.
In its modern all-order form it was first established in
\cite{Chetyrkin:1997un}.

Let us  remind  briefly the general setup  and current status   of the LET. 
We consider the (minimally) renormalized QCD with $n_h$ heavy and $n_l =n_f -n_h$ light
quarks. The corresponding Lagrangian reads\footnote{We omit not essential
  for  our reasonings
the  gauge-fixing and  ghost field containing terms.}
\be
\begin{split}
\ssL^{n_f}= -&\fr{1}{4}\,G^a_{\mu \nu} G^{a\,\mu \nu} 
  +\sum_{i=1, \dots,n_{\ell}}
  \ovl{\psi}_i  \lb\fr{i}{2}\overleftrightarrow{\slashed{\cal D}}-m_i\rb\psi_i 
+\sum_{i=(n_{\ell}+1),\dots, n_{h}}
  \ovl{\psi}_i  \lb\fr{i}{2}\overleftrightarrow{\slashed{\cal D}}-m_i\rb\psi_i 
  \\
  + & \la_0 H \Biggl(
  \sum_{i=(n_{\ell}+1), \dots, n_{f}} m_{i}\ovl{\psi}_i\psi_i
  +
 \sum_{i=1, \dots , n_{\ell}} m_i\ovl{\psi}_i\psi_i
  \Biggr)
  {}.
  \EQN{lag.nf}
\end{split}
\eeq
Here  the extended derivative and   the field strength tensor
are  
\beq \slashed{\cal  D} = \gamma_{\alpha} {\cal  D}_{\alpha},\,
{\cal  D}_{\alpha} = \partial_{\alpha} -i\,\, A^a_{\alpha} T^a,\,
%\eeq
%\beq
G_{\mu \nu}^{a}=
\p_\mu A_\nu^{a} - \p_\nu A_\mu^{a} 
+ g f^{abc} A_\mu^{b} A_\nu^{c}
{},
\eeq
with $g$ being quark-gluon coupling constant. In what follows
we will  also use  convenient combinations
\[
 O_1 \equiv -
\fr{1}{4}\,G^a_{\mu \nu} G^{a\,\mu \nu} {}, \ \  
\als = \frac{g^2}{4 \pi} \ \ \mbox{and} \ \  a=\frac{g^2}{4 \pi^2} \equiv \frac{\als}{ \pi}  
{}.
\]

% and the quark-gluon coupling constant $g$ is connected to 

For future reference we have
introduced into the QCD Lagrangian the Higgs-quark interaction. 
Having in mind various extensions of the SM containing either additional
quarks heavier than the top one or Higgs-like scalar particles with mass of
order a few GeV or even less \cite{Gorbunov:2023lga} we will consider a
generic case with the field\footnote{For simplicity we continue to refer to
  the field $H$ as  the Higgs one.} $H$ not necessarily being the one from
the SM.  Our only assumptions are: (i) the field $H$ couples with quarks via a
top-like Yukawa couplings as described in \re{lag.nf} and (ii)  its mass
$M_H$  \ice{larger than masses of light  quarks and less than}
meets the condition:
\beq m_i \gg M_H \gg m_j\ \ \mbox{ with} \   \  (n_\ell + 1) \le i \le n_f  \ \
\mbox{ and} \ 1 \le  j \le n_\ell {}.
\eeq
In the framework of the SM we naturally have $n_h=1$ and $n_\ell =5$.

Suppose that we consider some processes with the characteristic energy scale
$E$ below masses of all heavy quark masses. A typical  example of such a  process is 
the decay of the Higgs particle to {\em light} quarks and gluons. The
corresponding low-energy effective theory  is described by an effective
Lagrangian of the form \cite{Inami:1982xt}
(up to power-suppressed corrections of order $E/m_i$ or higher with 
$i \ge (n_\ell +1)$)
\ice{
\be
\begin{split}
\ssL^{n_l}= -&\fr{1}{4}\,\left(G^a_{\mu \nu} G^{a\,\mu \nu}\right)' 
  +\sum_{i=1, \dots , n_{\ell}}
  \ovl{\psi}'_i  \lb\fr{i}{2}\overleftrightarrow{\slashed{\cal D}}-m_i\rb\psi'_i 
  \\
  + & \la_0 H \Biggl(
   C_1\, O_1'
  +
  C_2\, O_2'\Biggr)
  {}.
  \EQN{lag.nl2}
\end{split}
{},
\eeq
}

\be
\ssL^{n_l}= -\fr{1}{4}\,\left(G^a_{\mu \nu} G^{a\,\mu \nu}\right)' 
  +\sum_{i=1, \dots, n_{\ell}}
  \ovl{\psi}'_i  \lb\fr{i}{2}\overleftrightarrow{\slashed{\cal D}}-m_i\rb\psi'_i 
    +  \la_0 H \Biggl(
   C_1\, O_1'
  +
  C_2\, O_2'\Biggr)
  {},
%  \EQNe{lag.nl}
 \EQN{lag.nl} 
 \eeq
 
\beq O_1' \equiv -
\fr{1}{4}\,\left(G^a_{\mu \nu} G^{a\,\mu \nu}\right)'{},\,\,
O_2'  =  \sum_{i=1, \dots ,  n_{\ell}} m'_i\, \ovl{\psi'}_i\psi'_i  
{}.
\eeq
Here   all primed quantities refer to QCD  c $n_\ell$ active quark  flavours;   
the effective quark-gluon coupling constant $g'$ and effective light
quark masses $m'_i$ are connected to the original ones via the corresponding {\em decoupling constants} (we
assume that one and the same $\msbar$ normalization parameter $\mu$ is
employed  in full and effective theories)
\beq
g'(\mu) = g(\mu)\, \zeta_{g}(\mu, \as(\mu), \unl{m}_h), 
\ \ \  m'_i(\mu) =   m_i \,\zeta_{m} \, (\mu, \as(\mu), \unl{m}_h)
\EQN{as_mq_dec}
{},
\eeq
where $\unl{m}_h = m_{(n_l+1)}, \dots m_{n_f} $ stands for heavy quarks
masses. Similar relations connect the fields in the effective Lagrangian with
the corresponding ones in the full one \re{lag.nf} (see,
e.g. \cite{Chetyrkin:1997un} for details). It  is also  convenient    to define
the decoupling constant for $\als$ and  $a$ as
\beq
\als'(\mu) = \als(\mu) \zeta_\alpha, \ \ a'(\mu) = a(\mu) \zeta_\alpha, \ \ 
 \zeta_\alpha :=  \zeta_g^2
{}.
 \EQN{zeta.alpha}
\eeq
The  all-order LET for the coefficient functions   $C_1$ and $C_2$ were found and  proven in
\cite{Chetyrkin:1997un}.
They read\footnote{ To be honest, LETs given below were established in
  \cite{Chetyrkin:1997un} only for a particular case of
  $n_h=1$. 
  However, a simple inspection of the derivation shows that it is easily
  extended to the general case with $n_h > 1$.  Note that a similar  observation has
  also been made in \cite{Grozin:2011nk}.}
\beq
C_1=
\sum_{h} m_h \ddp{m_h} \ln \zeta_\alpha
\equiv
\sum_{h} m_h \ddp{m_h} \ln \alpha_s'
\EQN{C1}
{}, \eeq
\beq
C_2= 1+
\sum_{h} m_h \ddp{m_h} \ln \zeta_m
\equiv
1+
\sum_{h} m_h \ddp{m_h} \ln m'_\ell
\EQN{C2}
{},
\eeq
where the index $h$ runs over all  heavy quarks (that is $h \in \{n_\ell+1,\dots n_f\}$) and
$m_l'$  stands for an  arbitrary light  quark mass in the effective  theory.
(We  have adapted prefactors to match our normalization of the operator $O_1$.)
It is worth noting that the above LETs hold  only  if both
full and  effective theories are minimally renormalized, including also quark masses.
For instance, they stop to be valid if quarks masses  are understood as   on-shell ones.

Let us  for the moment consider  the case of the SM with $n_h=1$ and $m_h = m_t$. Then, for example,
relation  \re{C1} can be explicitly written as:
\beq
C_1(\mu,a(\mu),m_t) = m_t \ddp{m_t} \ln \zeta_{\alpha} ( \mu,a(\mu),m_t) 
{}.
\eeq
Note that  on dimensional grounds $\zeta_{\alpha}$ depends only on ratio $\mu^2/m_t^2$. It opens the way
for extracting $(L+1)$-loop approximation for $C_1$ (and, obviously,  also for $C_2$) from
only $L$-loop result for $C_1$  via a use of standard renormalization group equations\footnote{
Provided $(L+1)$-loop  results for the QCD  $\beta$-function and the quark anomalous dimension
 are known beforehand.}.
For instance, the 3-loop results for the decoupling constants first obtained in
\cite{Chetyrkin:1997un} were used there (in combination with four-loop QCD RG functions
\cite{vanRitbergen:1997va,Czakon:2004bu,Chetyrkin:1997dh,Vermaseren:1997fq}) to
produce 4-loop predictions for $C_1$ and $C_2$.
The prediction for $C_1$
have been successfully tested against direct calculation only 20 years \mbox{later
\cite{Gerlach:2018hen}.}

At first glance it looks  that for $n_h >1$ the above described RG
considerations will stop to work.  In the next
section we will show that the this is not the case and extend the
``RG-improived'' versions of LETs for QCD with arbitrary many heavy quark
flavours. Then in Sections  3 and 4 we demonstrate how the new LETs help to
dramatically streamline some previous calculations and  even produce
completely  new  results.

Before  finishing  the Introduction  we  would  like to  note a  few things.
\begin{enumerate}
\item 
If $n_h >1$ but
  all masses of heavy quarks are considered to be equal 
  calculations of the decoupling constants are not more difficult than for a
  case of $n_h=1$. Thus, for us the condition $n_h >1$ will also mean that at
  least two heavy quark flavors have {\em different } masses. 
\item The  first calculation of the decoupling  constant at  two loops
  were done in \cite{Bernreuther:1981sg}.  Note that at  1- and 2-loops there is no technical differences
 between cases with $n_h=1$ and $n_h> 1$ (because all diagrams to compute contain  only one   
 quark flavour at a time).   
  The  calculation was  repeated (and slightly corrected)
  with a  different technique in \cite{Larin:1994va}. The  three loop calculations of \cite{Chetyrkin:1997un}
  have been checked and extended on  4-loop level in \cite{Chetyrkin:2005ia,Schroder:2005hy}. They all have been done
  under condition $n_h=1$.
\item The only work dealing with simultaneous decoupling of many heavy quarks
  is \cite{Grozin:2011nk}.  Here for the first time the authors have had to
  deal with 3-loop vacuum diagrams contributing to the decoupling constant and
  depending on {\em two} different quark masses. As for evaluation of $C_1$
  the contributing  3-loop vacuum diagrams were first computed in
  \cite{Anastasiou:2010bt}. 
   We will come  back to these  important works in Section 3.
\end{enumerate}

\section{RG-improved LETs  in QCD with $n_h>1$  \EQN{sec2}}

We start with some standard  RG  nomenclature.   We define the  QCD  beta function
and quark mass anomalous dimension as
%\vspace{-1mm}
\beq
\dmu  \, a = a\, \beta(a) = a \sum_{i >1} \beta_i a^i, \ \ \
\dmu \, m_i  = m_i\, \gamma_m(a) = m_i \sum_{i >1} (\gamma_m)_i a^i., \ \
\EQN{be.gm}
\eeq
Here $\dmu=\dmusq$, it is also convenient to define $\pmu=\pmusq$ and
$\p_{h} = \sum_h \, m_h \frac{\p}{\p m_h}$.
Note that the RG functions describe the RG evolution of the coupling constant
and the quark  masses  according  to  the  full QCD Lagrangian \re{lag.nf}.

For the case of the effective QCD \re{lag.nl} we will use the same
notations with added prime. That is, for example,
the effective $\beta'(a')$ is just a  shortcut for $\beta^{(n_l)}(a^{(n_l)})$
as well as $a' \equiv a^{(n_l)}$ and  so on. 

Let us consider the evolution equation for the decoupling
constant $\zeta_\alpha$.  Applying the operator $\dmu$ to
the relation
\beq
\ln a' = \ln(\zeta_\alpha) + \ln a 
{},
\eeq
we arrive to
\beq
\dmu \ln(\zeta_\alpha) = \beta'(a') -   \beta(a),
{}
\EQN{rg1}
\eeq
or, equivalently,
\beq
\left(\pmu +\gamma_m(a) \p_h
  +  \beta(a) a \frac{\p }{\p a} \right)
\ln \zeta_\alpha
= \beta'(a') -   \beta(a)
{}.
\EQN{rg2}
\eeq
Since elementary dimensional analysis implies 
\beq
(\p_{\mu^2} + \frac{1}{2} \,\p_h)\,  \ln \zeta_\alpha = 0
\EQN{rg3}
{},
\eeq
we combine \re{C1},(\ref{rg2}),(\ref{rg3})  and arrive to our final  RG-improved
LET for $C_1$
\beq
C_1 = -\frac{2}{1-2\, \gamma_m(a)}
\left(
  \beta'(a') -\beta(a) \,a \frac{\p \ln a'}{\p a}
\right)
{}.
\EQN{C1imp}
\eeq

The second RG-improved  LET is derived in the same way starting from applying
the  operator $\dmu$ to the  relation
\beq
\ln m' = \ln m +  \ln \zeta_m
{}.
\EQN{rg5}
\eeq
The resulting LET reads:
\beq
C_2 = 1 - \frac{2}{1-2\, \gamma_m(a)}
\left(
  \gamma'_m(a') -\gamma_m(a) - \beta(a)\, a\frac{\p \ln \zeta_m}{\p a}
\right)
{}.
\EQN{C2imp}
\eeq

A look on corresponding results in \cite{Chetyrkin:1997un}
(eqs. (41\textendash{}42))
prove the essential (up to normalization conventions) identity of RG-improved
LETs for $C_1$ and $C_2$ (see below) for the cases $n_h=1$ and $n_h > 1$.
The main advantage of LETs \re{C1imp} and \re{C2imp} over the previous ones
(see eqs.~(\ref{C1},\ref{C2})) is that the derivative with respect to the
coupling constant $a$ in the second terms in round brackets of
(\ref{C1imp},\ref{C2imp}) decreases by one the required loop order of the
decoupling constants $\zeta_\alpha$ and $\zeta_m$.  This will be used in the
next two sections.

\section{Example: $C_1$  at three loops in QCD with many  heavy  quarks}
 \ice{ 
1 - (4*a*lmm*tr)/3 + a^2*((32*ca*tr)/9 - (13*cf*tr)/3 + (16*lmm^2*tr^2)/9 + 
lmm*((-20*ca*tr)/3 + 4*cf*tr))
}

\newcommand{\Lmub}{L_{\mu }}

\newcommand{\Lmut}{\ln\frac{\mu^2}{m_t^2}}
\renewcommand{\Lmut}{L_{\mu t}}

\newcommand{\Lmum}{\sum_{h}\ln\frac{\mu^2}{m_h^2}}

\newcommand{\Lmuh}{L_{\mu h}}

\newcommand{\sigmah}{\sum_h \ln \frac{\mu^2}{m_h^2}}
\renewcommand{\sigmah}{L_{\mu h}}

We  start from  the old 2-loop result for $\zeta_\alpha$.
\be
\begin{split}
\zeta^{(6)}_\alpha = 1 & -  a^{(6)}(\mu)\,  \frac{ T_{F}}{3} \Lmut
\\
&+  a^{(6)}(\mu)^2 \, \left(
  \frac{2}{9} C_A T_F
  -
  \frac{13}{48} C_F T_F
  +(-\frac{5}{12} C_A  + \frac{C_F}{4})\,T_F \,\Lmut
  +\frac{T_F^2 \,\Lmut^2}{9}
  \right)
  {},
  \EQN{zeta6.2L}
\end{split}
\eeq
  where
  \beq
a^{(5)}(\mu) = \zeta^{(6)}_\alpha a^{(6)}(\mu)
{},
\EQN{dec6.2L}
\ee
$ \Lmut = \ln \frac{\mu^2}{m_t^2}$ and $m_t = m_t(\mu)$
stands for the (running) $\msbar$  mass of the top quark. 
$C_F$ and $ C_A$ are the quadratic Casimir
operators of the quark $[T^a T^a]_{ij} = C_F \delta_{ij}$ and the
adjoint $[C^a \, C^a]_{bd} = C_A \,  \delta_{bd}$, $(C^a)_{bc} = -if^{abc}$ 
representations of the Lie algebra.

Eq.~\re{zeta6.2L} was obtained for the standard QCD with $n_h=1$ and
$n_l=5$. It can be easily generalized to a general case with $n_h >1 $
due to the fact that at 1- and 2-loop level every diagram
contributing to \re{zeta6.2L} contains only {\em one} top quark loop.  Thus,
in order to extend \re{zeta6.2L} for the general case one needs to make three 
replacements ($L_{\mu h} = \sum_h \ln \frac{\mu^2}{m_h^2}$)
\[
  T_F^2 \Lmut^2 \rightarrow T_F^2 \sigmah^2, \  \
  T_F \Lmut \rightarrow T_F \sigmah, \ \
  T_F\rightarrow T_F \, n_h
{},
\]
with  the result
\be
\begin{split}
 \zeta_\alpha(a_s) & = 1 -  a(\mu)\,  \frac{T_{F}}{3} \Lmuh
\\
&+  a(\mu)^2 \, \left[
  \frac{2}{9} C_A T_F n_h
  -
  \frac{13}{48} C_F T_F n_h
  +\left(-\frac{-5}{12} C_A  + \frac{C_F}{4}\right)\,T_F \,\Lmuh
  +\frac{T_F^2}{9}  \,\Lmuh^2
  \right]
  {}.
  \EQN{zetanf.2L}
\end{split}
  \eeq

  Finally, a direct use the RG-improved LET \re{C1imp} and \re{zetanf.2L}
  leads to a general result for the coefficient function  $C_1$ at the 3-loop level\footnote{The
    necessary coefficients of the QCD $\beta$-function and the quark mass
    anomalous dimension $\gamma_m$ were first computed in
    \protect\cite{Tarasov:1980au,Tarasov:1982plg} and confirmed in
    \protect\cite{Larin:1993tp}.  }.
\beq
  C_1(a_s) =   a\, n_h \frac{2 T_{F}}{3}
  +  a^2 \, \left[
    \left(\frac{5}{6} C_A  -  C_F/2\right)\,T_F n_h
  - \frac{2}{9} T_F^2 n_h \,\Lmuh
  \right] + C_{1,3}\, a^ 3 +C_{1,4}\, a^4
  {},
  \EQN{C1.3l}
  \eeq
  where $a=a^{(nf)}(\mu)$,
%  \newpage
% input{eq3.5}
% my for -GG/4 in terms of as 
     \begin{eqnarray}									     
       C_{1,3}										     
       &&{=} \   										     
        C_F^2 T_F \frac{9}{16} n_h								     
      - C_F C_A T_F \left[ \frac{25}{18} n_h						     
      + \frac{11}{24} \Lmuh \right]  \nonumber\\						     
      &&{} - C_F T_F^2 \left[ \frac{5}{24} n_h n_l + \frac{17}{72} n_h^2			     
      - \Lmuh \left(\frac{1}{2} n_h  + \frac{1}{3} n_l \right) \right]			     
      + C_A^2 T_F \left[ \frac{1063}{864} n_h						     
      + \frac{7}{24} \Lmuh \right] \nonumber\\						     
      &&{}  - C_A T_F^2 \left[ \frac{47}{216} n_l - \frac{49}{432} n_h			     
      + \frac{5}{6} \Lmuh \right] n_h							     
      + \frac{2}{27} T_F^3 \Lmuh^2 n_h  							     
      \,.\EQN{C13.3l}									     
    \end{eqnarray}										   
and  the last term to the right of \re{C1.3l} is introduced for discussion in the
  next section.

  Expression\re{C13.3l} was first obtained via a direct (and  quite complicated)
  calculation of a large  number \ice{(at least 189)}  of 3-loop diagrams contributing to $C_1$
  including the ones depending on {\em two} different quark masses
  \cite{Anastasiou:2010bt}.
  It was later  confirmed in   \cite{Grozin:2011nk} (see the next Section).
%Our derivation is not  extemely simple but  reveals 

%\cite{Grozin:2011nk,Gorbunov:2023}
 
%\cite{Larin:1994va,Bernreuther:1981sg,Bernreuther:1983zp}

\section{New result: $C_1$ at  four  loops in QCD with many heavy  quarks}

\renewcommand{\text}[1]{#1}

The decoupling constants  $\zeta_\alpha$ and $\zeta_m$ have been analytically
computed for QCD with many heavy flavours at the 3-loop level in
\cite{Grozin:2011nk}.  The result includes complicated functions (di-logs,
etc.) of ratios $m_h/m_{h'}$. The authors of \cite{Grozin:2011nk} have used
the (RG-not-improved) LET \re{C1} and arrived essentially to the result
confirming \re{C13.3l} (and, naturally in agreement to
\cite{Anastasiou:2010bt}). They note: ``It is remarkable that although
$\zeta_\alpha$ contains di- and tri-logarithms there are only linear
logarithms present in $C_1$''. Our derivation of \re{C1.3l} presented above is
not only extremely straightforward but also reveals the reason behind this
remarkable simplicity: at the three loop level the coefficient function $C_1$ is contributed by
the 2-loop decoupling function only (not counting mass-independent $\beta$ and
$\gamma_m$).

But armed with the RG-improved LETs we can do more.  Indeed, as the four-loop
$\beta$ and $\gamma_m$ are known since long we could use LET \re{C1imp} in order to
upgrade the result \re{dec6.2L}  to one more loop\footnote{As we already have mentioned
for the case of $n_h=1$ it was done long ago in \cite{Chetyrkin:1997un}.}.

We start from some notations.  The result of \cite{Grozin:2011nk} for $\zeta_\alpha$ is convenient
to present as follows:
\beq \zeta_\alpha = 1+ d_{1} \,a + d_{2} \,a^2 + d_{3}\, a^3  
%d_{4} \, a^4
{}.
\eeq
Here, the coefficients $d_{1}$ and $d_{2}$ can be easily extracted from
% coefficients can be readily read off
\re{zetanf.2L}.  The
coefficient $d_3$ is a complicated function of $\mu$, $a(\mu)$ and quark mass
$m_{n_l+1}, \dots m_{n_f}$; the reader is referred to \cite{Grozin:2011nk} for
its full and complete description.

Finally, a use of RG-improved LET \re{C1imp} directly  leads for the following result
for  $C_1$ with the four-loop accuracy
\beq
C_{1}= -\frac{2}{1-2\, \gamma_m(a)}\left(
 \tilde{C}_{1,1} \,a + \tilde{C}_{1,2} \,a^2 + \tilde{C}_{1,3}\, a^3 + \tilde{C}_{1,4} \, a^4 
\right)
\EQN{C1.LET}
{},
\eeq
where
\be
\begin{split}
\tilde{C}_{1,1} =& -\beta_1 + \beta'_1,
\\
\tilde{C}_{1,2} =& -\beta_2 + \beta'_2 + (-\beta_1 + \beta'_1) \,  d_1,
\\
\tilde{C}_{1,3} =& -\beta_3 + \beta'_3 + (-\beta_2 + 2 \,  \beta'_2) \,  d_1 + \beta_1 \,  d_1^2 + (-2 \,  \beta_1 + \beta'_1) \,  d_2,
\\
\tilde{C}_{1,4} =& -\beta_4 + \beta'_4 + (\beta_2 + \beta'_2) \,  d_1^2 - \beta_1 \,  d_1^3
+ (-2 \,  \beta_2 + 2 \,  \beta'_2) \,  d_2
\\
&+
d_1 \,  (-\beta_3 + 3 \,  \beta'_3 + 3 \,  \beta_1 \,  d_2)
+ (-3 \,  \beta_1 + \beta'_1) \,  d_3.
\end{split}
\EQN{tildeC}
\eeq

We illustrate the result for $C_{1,4}$  with an example. Let us  consider
QCD with $n_f=6$ and $n_l=4$ (in other words we treat $t$- and $b$-quarks as  heavy
and $u,d,s$ and $c$-quarks as light ones).  
\ice{\beq
C_{1,4}=
\EQN{C14}
\eeq
}
Further, we expand $d_4$ in the limit $\frac{m_b}{m_t}  \to  0 $ and discard all
power-suppressed terms. The resulting coefficient
\[C^{{\scriptstyle as}}_{1,4} = C_{1,4} - \mbox{all power suppressed terms}\]
reads ($\Lmut = \ln\frac{\mu^2}{m_t^2}$,  $L_{\mu b} = \ln\frac{\mu^2}{m_b^2}$)
\ice{
C1asympas4 = -\frac{2079791}{62208} - {1}{324} L_{\mu t}^3 - \frac{1}{324} L_{\mu b}^3

+ (\frac{2299}{1728} - \frac{1}{108} L_{\mu b} )L_{\mu t}^2

-  \frac{6101}{5184}  L_{\mu b} + \frac{2299}{1728} L_{\mu b}^2

  +  (-\frac{1891}{1296} - \frac{1}{108} L_{\mu b}^2) L_{\mu t}  + \frac{491243}{13824} \zeta_3
}

\be
\begin{split}
C^{as}_{1,4}= -&\frac{2079791}{62208} - \frac{1}{324} L_{\mu t}^3 - \frac{1}{324} L_{\mu b}^3
 +  \left(\frac{2299}{1728} - \frac{1}{108} L_{\mu b} \right)L_{\mu t}^2
\\  -&  \frac{6101}{5184}  L_{\mu b} + \frac{2299}{1728} L_{\mu b}^2
+  \left(-\frac{1891}{1296} - \frac{1}{108} L_{\mu b}^2\right) L_{\mu t}
+ \frac{491243}{13824} \zeta_3
\end{split}
\EQN{C14}
\eeq
or, numerically:
\ice{
\be
\begin{split}
C^{as}_{1,4}= -&9.2829 - 0.00308642 L_{\mu t}^3 - 0.00308642 L_{\mu b}^3
 +  \left(1.33044 - 0.00925926 L_{\mu b} \right)L_{\mu t}^2
\\  -&  1.17689  L_{\mu b} + 1.33044 L_{\mu b}^2
 +  \left(-1.4591 - 0.00925926 L_{\mu b}^2\right) L_{\mu t} 
\end{split}
\EQN{C14N}
\eeq
}
\ice{
In[68]:= 33.4329 + 35.5355 1.20206

Out[68]= 76.1487
}
\be
\begin{split}
C^{as}_{1,4}= -&9.2829 - 0.00308642 \,L_{\mu t}^3 - 0.00308642 \,L_{\mu b}^3
 +  \left(1.33044 - 0.00925926 \,L_{\mu b} \right)\,L_{\mu t}^2
\\  -&  1.17689  \,L_{\mu b} + 1.33044 \,\,L_{\mu b}^2
+  \left(-1.4591 - 0.00925926 \,L_{\mu b}^2\right) \,L_{\mu t}
{}.
\EQN{C14N}
\end{split}
%\EQN{C14N}
\eeq
It is clear that potentially large logarithmic terms in \re{C14N} should be
possible to sum with the help of the renormalization group technique like it
was done long ago in \cite{Inami:1982xt} for the case of a single heavy
quark. We are going to consider the issue in future.

\section{Conclusion}

The main result of the present work is the extension of two LETs describing
effective coupling of the Higgs field with gluons and light quarks
respectively for the case of QCD with many heavy quarks.

We have demonstrated the RG-improved LET for the coefficient $C_1$ allows to
extend the corresponding result of \cite{Grozin:2011nk} by one more loop.
We have also reproduced the 3-loop result for $C_1$ in QCD with many heavy
quarks.

We would like to stress that both results for $C_{1,3}$ and $C_{1,4}$ have
been obtained in a very simple way without any extra complicated calculations
(of course, we have heavily used the old results for the decoupling constant
$\zeta_{\alpha}$ in 2- and 3-loop approximations as found in
\cite{Bernreuther:1981sg,Larin:1994va} and \cite{Grozin:2011nk} respectively).

The author is very grateful to Sasha Pukhov who ignited his interest to the
problem of finding the coefficient function $C_1$ in QCD with many heavy
quarks. He also thanks  the referee for careful reading  the manuscript. 

%\end{document}

%%%%%%%%%%%%%%%%%%%%%%%%%%%%%%%%%%%%%%%%%%%%%%%%%%%%%%%%%%%%% 
\providecommand{\href}[2]{#2}\begingroup\raggedright\endgroup

\end{document}